# Quantum transport in electron Fabry-Perot interferometers

F. E. Camino, Wei Zhou, and V. J. Goldman

*Department of Physics, Stony Brook University, Stony Brook, NY 11794-3800, USA*

We report experiments on Fabry-Perot electron interferometers in the integer quantum Hall regime. The GaAs/AlGaAs heterostructure devices consist of two constrictions defined by etch trenches in 2D electron layer, enclosing an approximately circular island. The interferometer is formed by counterpropagating chiral edge channels coupled by tunneling in the two constrictions. Interference fringes are observed as conductance oscillations, similar to the Aharonov-Bohm effect. Front gates deposited in etch trenches allow to fine-tune the device and to change the constriction filling $f$ relative to the bulk filling. Quantum-coherent conductance oscillations are observed on the $f = 1 - 4$ plateaus. On plateau $f$ we observe $f$ conductance oscillations per fundamental flux period $h/e$. This is attributed to the dominance of the electron-electron Coulomb interaction, effectively mixing Landau level occupation. On the other hand, the back-gate charge period is the same (one electron) on all plateaus, independent of filling. This is attributed to the self-consistent electrostatics in the large electron island. We also report dependence of the oscillation period on front-gate voltage for $f = 1, 2$ and 4 for three devices. We find a linear dependence, with the slope inversely proportional to $f$ for $f = 1$ and 2.

## I. INTRODUCTION

The integer quantum Hall effect[1] can be understood in terms of transport by edge channels corresponding to an integer number of fully occupied Landau levels.[2-4] In this picture, near an integral Landau level filling $\nu \approx f$, when the chemical potential lies in the gap of localized bulk states, the current is carried by dissipationless edge channels and the Hall resistance is quantized to $h/fe^2$. Dissipative transport occurs when current is carried either by extended bulk states of the partially occupied topmost Landau level, between the plateaus, or by quantum tunneling between the extended edge states. Such interpretation of the IQHE of non-interacting electrons in terms of edge channels is straightforward since for non-interacting electrons the edge channels are formed in one-to-one correspondence with the bulk Landau levels defined in the single-electron density of states. However, as is well known, the electron - electron interaction is not small compared to single-particle energies involved, and the effects of interaction are subjects of intense experimental and theoretical research.

In this paper we present a detailed experimental characterization of electron Fabry-Perot interferometers in the integer quantum Hall (QH) regime. These studies are motivated in part by application of the same interferometer devices in the fractional QH regime, where interference of fractionally-charged Laughlin quasiparticles has been studied.[5-9] Similar electron interferometer devices have been studied by others in the integer QH regime.[10-12] Additional motivation is provided by proposed application of Fabry-Perot interferometers, in conjunction with quantum antidots, to detection of non-Abelian braiding statistics,[13-20] thus verifying the microscopic ground state of certain observed fractional QH states, such as the even-denominator $f = 5/2$.

The Fabry-Perot interferometers have geometry complementary to quantum antidots, small potential hills lithographically defined in the 2D electron plane.[21-23] In a quantizing magnetic field $B$, in QH regime, the electron states bound on the antidot display discrete energy spectrum; these quantized states are probed by resonant tunneling,[24-28] enabled by placing the quantum antidot in a constriction. Experiments on quantum antidots show that the fundamental magnetic flux period $h/e$ consists of $f$ more or less equally spaced tunneling conductance peaks, where $f$ is the QH filling of the constriction plateau where the peaks are observed. In the integer QH regime this can be understood as following from the fact that on such plateau $f$ Landau levels are occupied.[27,28] As we report in this work, many features of interferometric conductance oscillations are similar to the resonant tunneling in quantum antidots. Although dynamically different, the stationary state structure in both types of devices is determined by the Aharonov-Bohm quantization in an interacting 2D electron system.

## II. SAMPLES AND EXPERIMENTAL TECHNIQUES

The interferometer samples were fabricated from very low disorder modulation-doped GaAs/AlGaAs heterostructures. The 2D electron system is buried 240 – 320 nm below the surface. First, Ohmic contacts are formed on a pre-etched mesa. Then etch trenches are defined by electron-beam lithography, using proximity correction software for better definition of narrow and long gaps between the exposed areas. After a shallow 120 – 180 nm wet etch, 50 nm thick Au/Ti front-gate metallization is deposited in a self-aligned process. Finally, samples are mounted on sapphire substrates with In metal, which serves as the global back gate.

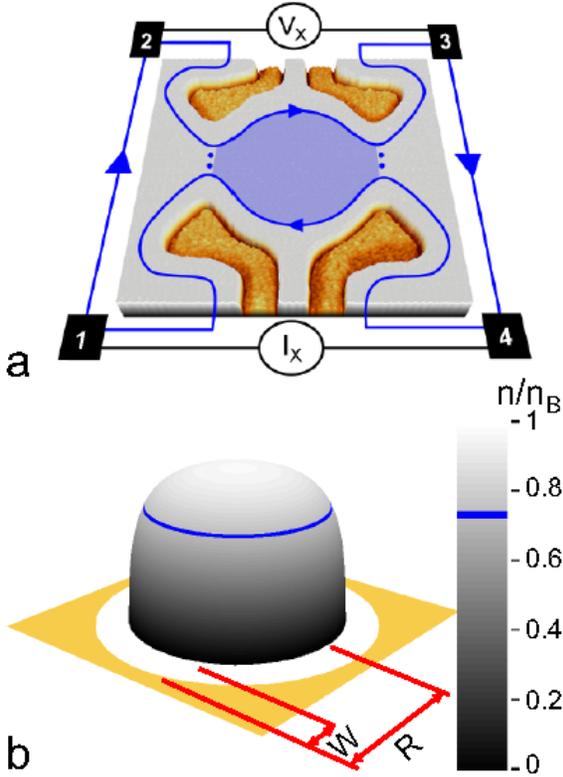

FIG. 1. A Fabry-Perot electron interferometer device. (a) Topography of the interferometer region obtained with an atomic force microscope. Four front gates are deposited in shallow etch trenches, defining a circular island separated from the 2D bulk by two wide constrictions. The lithographic radius of the island $R \approx 1.05$ μm. The chiral edge channels (blue lines) follow an equipotential at the periphery of the undepleted 2D electrons. Tunneling (blue dots) occurs at the saddle points in the constrictions. The edge channel path is closed by the tunneling links, thus forming the interferometer. Ohmic contacts (four numbered pads) are located at the corners of the 4×4 mm sample. The back gate (not shown) extends over the entire sample; it is separated from the 2D electron plane by a 0.43 mm GaAs substrate. (b) Illustration of electron density profile in a circular island. Note that constrictions are ignored in this model.[29,30] The mesa etch (yellow) creates a depleted region of width $W \approx 245$ nm (white annulus). The radius of the blue ring $r \approx 680$ nm is determined from the experimental Aharonov-Bohm period.

Samples were cooled in the tail of the mixing chamber of a top-loading into mixture dilution $^3$He-$^4$He refrigerator. A bulk 2D electron density $n_B = 0.9 \cdots 1.25 \times 10^{11}$ cm$^{-2}$ was achieved after illumination by a red LED at 4.2 K. All experiments reported in this work were performed at the fixed bath temperature of 10 mK, calibrated by nuclear orientation thermometry. Extensive cold filtering in the electrical leads attenuates the electromagnetic background "noise" incident on a sample, allowing to achieve effective electron temperatures of ≤15 mK.[8] Four-terminal longitudinal $R_{XX} = V_X / I_X$ and Hall $R_{XY} = V_Y / I_X$ magnetoresistances, see Fig. 1(a), were measured with a lock-in technique at 5.4 Hz. The excitation current was set so as to keep the larger, Hall or longitudinal voltage ≤5 μV.

## III. RESULTS AND DISCUSSION
### A. Magnetotransport

Even at zero front-gate $V_{FG} = 0$, the GaAs surface depletion of the etch trenches, which remove the doping layer, creates electron confining potential. The etch trenches define two ≈1.2 μm lithographic width constrictions, which separate an approximately circular electron island from the 2D electron "bulk". The $B = 0$ shape of the electron density profile in a circular island resulting from mesa depletion[29,30] is illustrated in Fig. 1(b). In these large islands with 2 - 4×10³ electrons, the 2D electron density profile is determined mostly by the classical electrostatics, minimizing the energy of electron-electron repulsion, compensated by interaction with the positively charged donors. The real interferometer device depletion potential has saddle points in the constrictions, and so has the resulting density profile. In a quantizing magnetic field edge channels form, but the overall electron density profile closely follows the $B = 0$ profile in these relatively large devices so as to minimize total Coulomb energy.

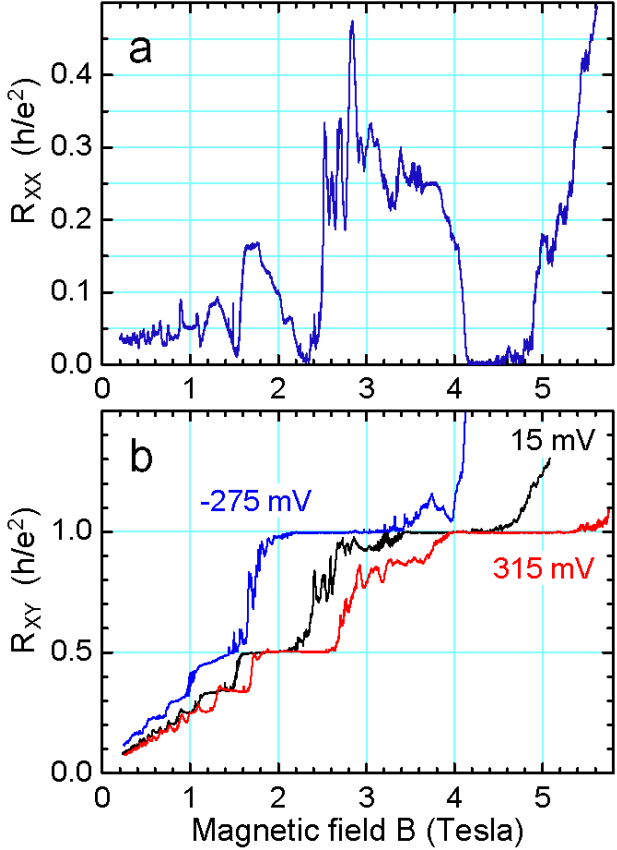

FIG. 2. (a) Longitudinal magnetoresistance of the interferometer at front-gate $V_{FG} \approx 0$. (b) Hall resistance data at three front-gate voltages, shown in labels. Application of $V_{FG}$ changes the electron density in the interferometer region, both the island and the constrictions, thus shifting the $B$-positions of the quantized plateaus. The smallest filling factor, that in constrictions, determines the Hall signal, while the longitudinal signal depends on filling in all regions of the sample, including the 2D bulk.

Figure 2 shows four-terminal magnetoresistances in sample M97Bm. Because in a uniform $B$ the Landau level filling factor $\nu = hn/eB$ is proportional to local electron density, in the depleted regions of the sample $\nu$ is different from the 2D bulk $\nu_B$. While $\nu \propto n/B$ is a variable, the quantum Hall exact filling $f$ is a quantum number defined by the quantized Hall resistance as $f = h/e^2 R_{XY}$. Because QH plateaus have finite width, regions with a bit different $\nu$ may have the same $f$. In samples with lithographic constrictions, in general, there are two possibilities: (i) when depletion is small, the whole sample may have the same QH filling $f$; and (ii) more often, the constriction filling $f$ and the bulk filling $f_B$ are different.

The Hall resistance $R_{XY}$ is determined by the filling in the constrictions, its plateau positions in $B$ giving definitive values of $f$. Thus Fig. 2(b) clearly shows transistor action of the front gates. As discussed below (and reported before[7,30]), the position of conductance oscillations in $B$ experiences similar shift with $V_{FG}$: the $B$-range where oscillations occur corresponds to a specific $f$. The longitudinal $R_{XX}$ has QH minima and quantized plateaus at $R_{XX} = (h/e^2)(1/f - 1/f_B)$, when plateaus in constrictions and the bulk happen to overlap in $B$.

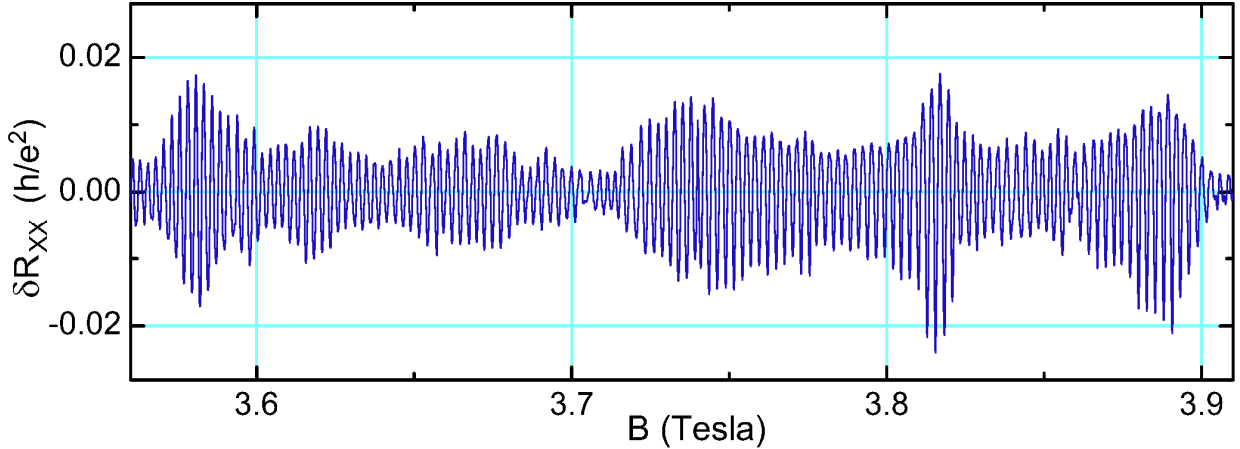

FIG. 3. Representative Aharonov-Bohm oscillatory signal on $f = 1$ plateau. Up to 250 periods can be seen in a full trace.

The fine structure in the traces of Fig. 2 is caused by tunneling and quantum interference effects, including Aharonov-Bohm-like conductance oscillations. Fig. 3 shows blow-up of a typical oscillatory signal $\delta R_{XX}$, obtained by subtracting a smooth background from the directly measured $R_{XX}$. Corresponding oscillations are also observed in the Hall $R_{XY}$. The origin of the conductance oscillations is the $\propto \cos(2\pi \Phi e/h)$ modulation term in the tunneling conductance, where $\Phi$ is the magnetic flux through the closed electron orbit encircling the island.

**B. Dependence of oscillation period on filling**

Aharonov-Bohm interference occurs when electron tunneling at the saddle points in the constrictions connects the counterpropagating edge channels, Fig. 1(a). Tunneling completes a quantum-coherent closed path that encloses a well-defined area $S$ in the 2D electron island. Because tunneling amplitude becomes exponentially small when tunneling distance exceeds several magnetic lengths, the region where measurable tunneling occurs is nearly fixed at the saddle points in the constrictions. Thus, when the oscillatory interference signal is observed, the electron density and filling of the island closed path (an equipotential) is determined by the saddle point in the constrictions.

The oscillatory conductance $\delta G$ is calculated from the directly measured $\delta R_{XX}$ and the quantized Hall resistance $R_{XY} = h/fe^2$ as $\delta G = \delta R_{XX}/R_{XY}^2$, a good approximation for weak tunneling, $\delta R_{XX} \ll R_{XY}$. Figure 4 shows oscillatory conductance for the $f = 1$, 2 and 4 constriction plateaus. The corresponding magnetic field periods are $\Delta_B = 2.71$, 1.38 and 0.67 mT, respectively. We notice that $B$-periods scale with QH filling so that the product $f \Delta_B \approx 2.7$ mT is approximately constant. The Aharonov-Bohm oscillations are periodic with magnetic flux enclosed by the electron path, the fundamental period being $h/e$. Thus, expecting that the Aharonov-Bohm edge channel area $S$ does not change much with $f$ in these large devices, on $f$-th plateau we see $f$ oscillations per flux period $\Delta_\Phi = h/e$. As discussed below, this is not what is expected for non-interacting electrons in the integer QH regime.

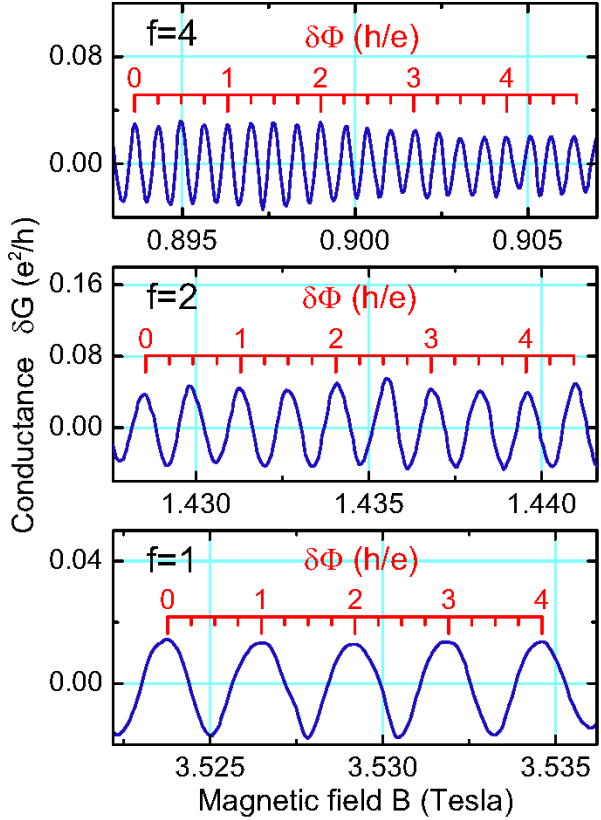

FIG. 4. Oscillatory conductance on $f = 1$, 2 and 4 constriction plateaus. All three panels have the same width in magnetic field. On the $f$-th plateau, there are $f$ oscillations per fundamental flux period $\Delta_\Phi = h/e$.

A complementary experiment is to vary the *back-gate* voltage $V_{BG}$ while keeping $B$ fixed. Figure 5 shows thus obtained oscillatory conductance data. The back-gate periods are $\Delta_{V_{BG}} = 364$, 358 and 412 mV on the $f=1$, 2 and 4 constriction plateaus, respectively. The relative difference between $f=1$ and 2 periods is ~2%, while between $f=1$ and 4 periods is ~12%. We conclude that back-gate periods $\Delta_{V_{BG}}$ are nearly constant, independent of $f$ in these experiments. As discussed below, they correspond to addition of one electron to the area enclosed by the interference path.

Figure 6 summarizes the experimental oscillation periods obtained from the $B$- and $V_{BG}$-sweep data, as shown in Figs. 4 and 5. The fits confirm the approximate relations $\Delta_B \propto 1/f$ and $\Delta_{V_{BG}} = const$. Similar relations for the periods were also reported for quantum antidot samples.[27,28] These results can be understood as follows. Although dynamically different, the stationary state structure in both types of devices is determined by the Aharonov-Bohm quantization in an interacting 2D electron system. The electron island in these devices is large, containing several thousand electrons. The main effect of the confinement potential is to lift the massive degeneracy of the single-electron states in each Landau level. Each conduction oscillation corresponds to a change by one in the number of the electron states in $S_\mu$, the area enclosed by the electron orbital at the chemical potential $\mu$.[7,28] As a function of $B$, when $S_\mu$ is nearly fixed by the confining potential, we expect one oscillation per filled Landau level when the flux through $S_\mu$ is changed by $h/e$, that is, $f$ oscillations (if resolved) per fundamental flux period $\Delta_\Phi = h/e$.

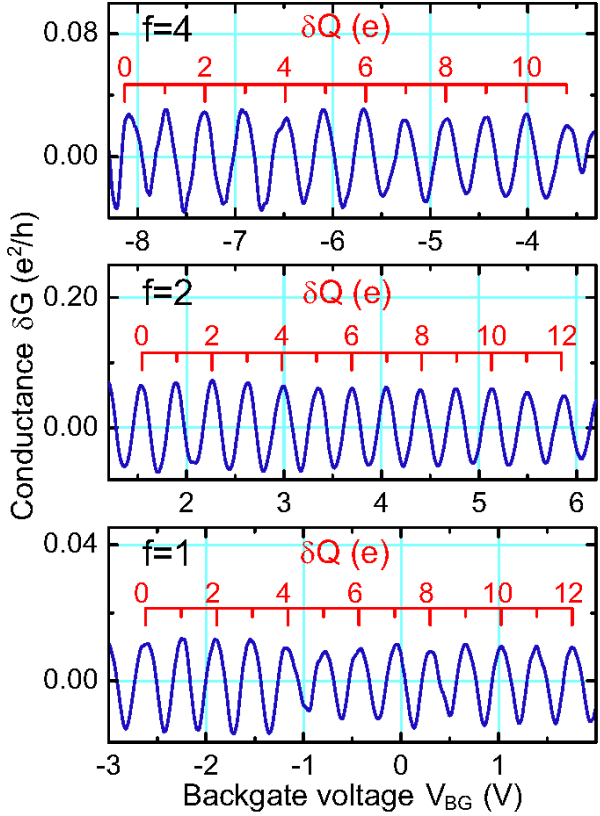

FIG. 5. Oscillatory conductance vs. back-gate voltage on $f=1$, 2 and 4 constriction plateaus. All three panels have the same width in voltage. On the $f$-th plateau, the charge period $\Delta_Q \approx e$, independent of $f$.

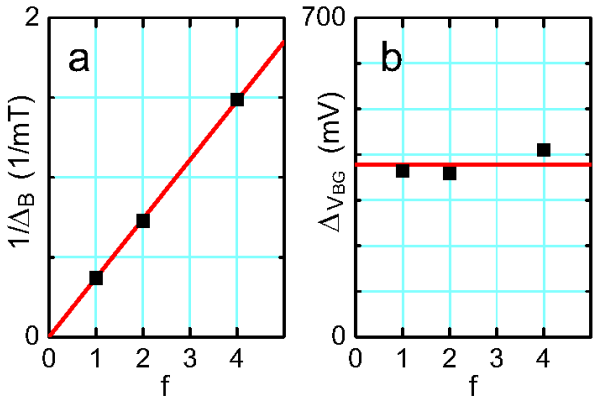

FIG. 6. Summary of magnetic field (a) and back-gate voltage (b) periods on $f=1$, 2 and 4 constriction plateaus. The lines show the linear fits $\Delta_B \propto 1/f$ and $\Delta_{V_{BG}} = const$.

For non-interacting electrons, in each spin-polarized Landau level the single-electron states are quantized by the Aharonov-Bohm condition: flux through the area of an encircling orbital satisfies $\Phi = BS_m = m(h/e)$, where $m = 0, 1, 2, \ldots$ is the quantum number of the orbital.[2-4] Thus, $S_m = m(h/eB) = 2\pi m \ell^2$, where the magnetic length $\ell = \sqrt{\hbar/eB}$, and $S_{m+1} - S_m = h/eB = 2\pi \ell^2$ is the area per single-electron state. The main effect of the confinement potential is to lift the massive degeneracy of single-electron states in each Landau level. In the first order perturbation theory, the quantization $S_m = 2\pi m \ell^2$ is not affected by confinement, the confinement potential is simply added to the cyclotron and spin energies. Thus, on the $f = 1$ QH plateau, each conductance oscillation corresponds to a change by one in the number of electron states within $S_\mu$. When $S_\mu$ is nearly fixed by the confinement potential, the flux period is one $h/e$, and the interference path area can be determined from the field period, $S_\mu = h/e\Delta_B$.

For interacting electrons, when the occupation of particular Landau levels is mixed, the island electron states are superpositions of the basis orbitals in different Landau levels. The sum rules apply however, specifically, on $f$-th QH plateau there are $f$ electron states per area enclosing flux $h/e$. When flux through area $S_\mu$ is increased by one more $h/e$, there are $f$ island electron states crossing $\mu$. Thus, we expect $S_\mu \approx h/ef\Delta_B$, and $f$ conductance oscillations per fundamental flux period $h/e$, provided all the oscillations are resolved. This conclusion is supported by theoretical models considering on-site (within the island) Coulomb interaction.[31-35]

Changing magnetic field does not affect the equilibrium electron density in the large island, but redistributes the electron occupation between Landau levels. Application of a positive back-gate voltage $V_{BG}$ increases electron density. In a fixed $B$, when the density of states in each Landau level in an area is fixed too, application of $V_{BG}$ changes occupation of these states. Because the back gate is remote, its effect is a small perturbation: $V_{BG} = 1$ V changes electron density by $0.0016 n_B$. Because the back gate is global, extending over the entire sample, its effect, to the first order, does not change the island confinement potential (unlike the front gates), so that $S_\mu$ remains nearly constant. Thus, we interpret the back-gate oscillations as due to addition of electrons: one oscillation with period $\Delta_{V_{BG}}$ corresponding to addition of one electron to the island area $S_\mu$.

Similar to quantum antidots,[21,27] the 2D electron charge density attracted by the back gate can be deduced from the parallel-plate capacitor formula. The quantum corrections[25] are on the order of $10^{-5}$. The net charge variation $\delta Q$ within the interference area $S_\mu$ is proportional to back-gate voltage, $\delta Q = \alpha(\delta V_{BG}/f\Delta_B)$. This relation normalizes the back-gate voltage periods by the experimental $B$-periods, approximately canceling the difference in device area for different samples and due to a front-gate bias. The coefficient $\alpha$ is known *a priori* in quantum antidots to a good accuracy because the antidot is completely surrounded by a quantum Hall fluid. In an interferometer, however, the island is separated from the 2D electron plane by the depleted front-gate etch trenches, so that its electron density is not expected to increase by precisely the same amount as the bulk $n_B$.

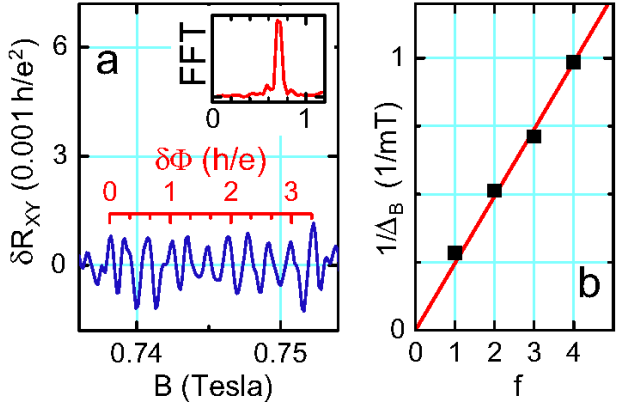

FIG. 7. (a) Weak conductance oscillations observed for $f = 3$ constriction plateau ($V_{FG} = -275$ mV). Fast Fourier transform (FFT) gives the period $\Delta_B = 1/(0.71 \text{ mT}^{-1}) = 1.4$ mT. (b) The same Hall $R_{XY}$ vs. $B$ trace also displays $f = 1$, 2 and 4 oscillations; the inverse $B$-period is plotted vs. constriction filling.

Conductance oscillations for the $f = 3$ constriction plateau are more difficult to observe since this plateau has a relatively small gap, being in the second spin-split Landau level. Evidence for such oscillations, Fig. 7(a), has so far been seen in only one experimental trace, which also displays strong $f = 1$, 2 and 4 oscillations. All four magnetic field periods $\Delta_B$ from this single trace obey the scaling relation presented above, $\Delta_B \propto 1/f$, as illustrated in Fig. 7(b).

### C. Dependence of oscillation period on front-gate bias

Application of a front-gate voltage $V_{FG}$ appreciably affects the island confining potential.[30] It also has a transistor effect, changing the overall island electron density. The effect of $V_{FG}$ on the constriction density, see Fig. 2, was discussed above. Because the electron interference path follows an equipotential passing near the saddle points in constrictions, the area $S_\mu$ is affected too. This is evidenced by the change of the measured oscillation periods $\Delta_B$ upon application of $V_{FG}$.[7,30] Figure 8 summarizes the $\Delta_B$ vs. $V_{FG}$ data for three samples: M97Bm reported in this work and in Refs. 5 - 8, M61Dd reported in Refs. 7 and 30, and M97Ce in Ref. 9, including some unpublished data. The front-gate voltage dependencies for different samples and for different cooldowns of the same sample are scaled appropriately so as to give equal $\Delta_B(V_{FG} = 0)$ at $f = 1$ constriction plateau.[7]

The raw (unscaled) data show an approximately linear dependence $\Delta_B(V_{FG})$ in the limited range of $V_{FG}$ studied. This is expected because the main confinement is provided by the etch trenches, not the front gates. Likewise, an approximately linear dependence is also obtained for the inverse $S_\mu = h/e\Delta_B$ vs. $V_{FG}$ dependence.[30] The two dependencies differ by a small term, quadratic in $V_{FG}$, which is not much more than the experimental error and thus can not be determined reliably. The linear fits yield the slopes, $d\Delta_B/dV_{FG}$, for each constriction filling. The $f = 1$ and 2 slopes scale as $f(d\Delta_B/dV_{FG}) \approx const$; this relation is in agreement with the scaling relation derived in Ref. 7. The $f = 4$ slope does not fit this relation well, however, perhaps due to the large experimental uncertainty, or because the $f = 4$ oscillations occur at low $B < 1$ T, where the simple fixed edge channel picture is not a good approximation. We note that the back-gate oscillation period data for $f = 4$, Fig. 5, is also a bit off from that for $f = 1$ and 2.

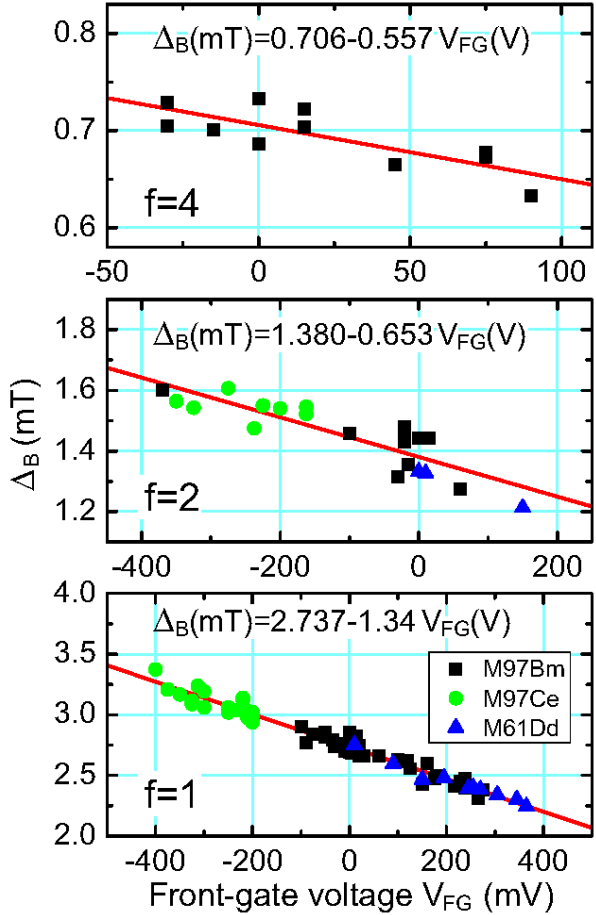

FIG. 8. Dependence of the magnetic field period $\Delta_B$ on front-gate voltage for $f=1$, 2 and 4 constriction plateaus. Data from three samples are summarized. Front-gate data are scaled so as to make $\Delta_B(V_{FG}=0)$ to coincide. The lines show the linear fits $\Delta_B = a - bV_{FG}$, the fit parameters are given in labels.

That a negative front-gate voltage should decrease the interference area $S_\mu$ is not obvious *a priori*. In addition to decreasing the overall electron density in the interferometer region, the front gates modify the island and the constriction electron density profile by affecting the primary confining potential of the etch trenches.[7] Since tunneling amplitude is exponentially sensitive to the tunneling distance, the position of the tunneling links at the saddle points in the constrictions is nearly fixed. The constrictions' saddle point electron density determines the equipotential contour of the Aharonov-Bohm path in the island. As evidenced by the systematic increase of the period $\Delta_B$ (decrease of island area) with negative $V_{FG}$, the saddle point electron density decreases proportionately less than the island center density. This experimental result is counterintuitive if one neglects the fact that the front gates have long leads and surround the island, while being only to one side of a constriction. Accordingly, the island edge channels must follow the constant electron density contours with density equal that in the constrictions and move inward, towards the island center, and the interference path area shrinks. The electronic charge $Q$ within $S_\mu$ decreases because both: the overall island density decreases, and also because the area itself decreases.

## IV. CONCLUSIONS

In conclusion, we have experimentally studied quantum electron transport in Fabry-Perot interferometers in the integer quantum Hall regime. In these two-constriction devices, electrons execute a closed path around a large 2D electron island. Both the magnetic field $\Delta_B$ and the back-gate charging $\Delta_{V_{BG}}$ periods correspond to excitation of one electron per oscillation within the island QH fluid enclosed by the interference path. On the island QH plateaus $f = 1 - 4$, we find that $\Delta_B \propto 1/f$, so that the fundamental flux period $\Delta_\Phi = h/e$ contains $f$ oscillations. This is interpreted as evidence of the dominance of the electron-electron interaction in the island, which mixes the single-particle Landau level occupation. We also present the dependence of the oscillation periods on front-gate voltage $V_{FG}$ in three interferometer devices. These data support a counterintuitive conclusion that the constriction electron density is affected by the front gates less than that in the island center.

We acknowledge discussions with D. V. Averin, B. I. Halperin and B. Rosenow. This work was supported in part by the NSF DMR.